\documentclass[manuscript,nonacm]{acmart}

\AtBeginDocument{%
  }

\setcopyright{acmlicensed}
\copyrightyear{2026}
\acmYear{2026}
\acmDOI{XXXXXXX.XXXXXXX}
\acmConference[ACM AI Letters]{ACM AI Letters}{2026}{Woodstock, NY}

\begin{document}

\title{Why AI Slop Matters, but Not Like That}

\author{Sachita Nishal}
\email{nishal@u.northwestern.edu}
\orcid{0000-0001-6192-6091}
\affiliation{%
  \institution{Northwestern University}
  \city{Evanston}
  \country{USA}
}

\author{Marijn Sax}
\email{m.sax@uva.nl}
\orcid{0000-0002-9980-9154}
\affiliation{%
  \institution{University of Amsterdam}
  \city{Amsterdam}
  \country{Netherlands}
}
\author{Kimon Kieslich}
\email{k.kieslich@uni-hohenheim.de}
\orcid{0000-0002-6305-2997}
\affiliation{%
  \institution{University of Hohenheim}
  \city{Stuttgart}
  \country{Germany}
}

\affiliation{%
  \institution{University of Amsterdam}
  \city{Amsterdam}
  \country{Netherlands}
}

\renewcommand{\shortauthors}{Nishal, Sax, \& Kieslich}

\begin{abstract}
This is a response to the paper ``Why Slop Matters''. By offering both immanent and external critique, we argue that the authors' reasoning neglects the socio-technical context of AI slop. Our paper presents an ethical and social science informed response that centers the debate on the social function and aesthetic value of AI slop. We conclude that AI slop is an important research subject but call for a contextual and culturally-grounded debate on the issue. To that end, we discuss some key elements of an agenda for future research on the phenomenon of AI slop.
\end{abstract}



\maketitle

\section{Introduction}

In their provocative article `Why Slop Matters,' \citet{kommers2025slop} argue that contrary to the common idea that AI-generated slop is a form of pollution, it serves a social function and has its own aesthetic value. 
We share their conviction that AI slop warrants rigorous study, but we argue that their framework, which is based on individual preference satisfaction, forecloses the social and political aspects of AI slop that matter most.
To that end, in this piece, we deconstruct their argument through immanent and external critique, and propose a more critical research agenda for studying AI slop. \looseness -1

The satisfaction of individual desires through AI-generated, hyper-personalized content is a pervasive frame in both industry\footnote{For an example, see Mark Zuckerberg's note titled \textit{\href{https://www.meta.com/superintelligence/}{Building Personal Superintelligence}}, in which he describes the goal of AI as enabling individuals ``to direct it towards what they value in their own lives''.} and academic discourses on generative AI.
Kommers et al.'s analysis exemplifies this approach too, since their argument for slop's social function asks whether it satisfies users' preferences (the kid-father example), and their aesthetic argument asks whether consumers might come to value it (the kitsch analogy).
This frame carries implicit assumptions about the nature of people's preferences, AI's (social) costs, and aesthetic value, and thus it shapes what questions can and cannot be asked about AI slop.

For one, people are not always preference-satisfying consumers with static, self-derived choices who encounter content and proceed to then accept or reject it. 
They are also participants in communities through which preferences are formed, identities constituted, and meanings made collectively \cite{senDevelopmentFreedom1999}.
Rather than containing our inquiry into AI slop to whether it satisfies individual preferences, we can broaden to ask what kinds of communities, relationships, and conditions of meaning-making it assembles or forecloses, and what kinds of impacts it has on them.

Because AI-generated slop has the potential to radically change---or: pollute?---the online environments where people post, share, and consume all kinds of information, we believe the terms on which we study and evaluate it matter greatly.
The rest of this work offers our sociotechnical critiques and research agenda for studying AI slop, through which we hope to broaden the criteria through which the value of AI slop is evaluated.

\section{The Arguments for Slop and Its Internal Inconsistencies}

Kommers et al. define AI slop with three prototypical properties: \textit{(1) Superficial Competence}. AI slop demonstrates a kind of competence that would take real expertise or skill to achieve without AI; this is belied by a lack of underlying substance, craft, or communicative intent. \textit{(2) Asymmetric Effort}. AI slop is generated with a facile prompt and requires little or no effort of the kind that would be necessary to create such an output without AI. \textit{(3) Mass Producibility}. AI slop is designed to work within a digital ecosystem of production and distribution to reach a mass audience.  

From this, their argument for the ``social function of slop'' goes like this: people have weirdly idiosyncratic preferences they want to satisfy when searching for content. 
As a collective, humanity can produce astronomical amounts of content, meaning that there is no supply-side problem in terms of \textit{quantity}. 
There is, however, a supply-side  problem in terms of \textit{quality}: with such a plethora of people holding so many different idiosyncratic preferences, producing content that meets each person's specific desires is difficult at scale.
This is where their core example comes in: 

\begin{quote}

``For example, many children would find it entertaining to watch a video of their father dancing wildly in his underwear. The entertainment value comes mainly from the fact that it is one's own father—not someone else's. Crucially, there is an insuperable cost to generating this content under usual circumstances: getting your dad to do the dance. AI Slop can eliminate this bottleneck.''

\end{quote}

This example is meant to show that AI slop can satisfy particular idiosyncratic preferences, and hence offers a social function. 
The video of the dancing father would \textit{only} be interesting to the child in question, and without generative AI, the video probably would never exist. 

Second, the argument for the aesthetic value of AI slop takes the form of a hypothetical: what if AI slop did, somehow, turn out to have aesthetic value because there are cyclical trends to be discovered on the demarcation between `low' art and `real' art? 
They refer to the history of kitsch, which was originally evaluated as cheap, formulaic, mechanical content, but was later---at least by some---reevaluated ``as having obvious cultural value'' \cite{kommers2025slop}. 
The argument, as we understand it, is that if AI slop is sufficiently similar to `low' culture, and if `low' culture has been rehabilitated in the past, then AI slop may also be rehabilitated in the future.

Each of these arguments presents internal inconsistencies in light of how AI slop is defined, meaning that either the definition does not properly capture the phenomenon of AI slop, or the arguments do not engage with the phenomenon of AI slop. 
We break them down below.

The example of the kid generating a video of their father describes content created for private consumption based on a specific, idiosyncratic preference, but this runs counter to the definition of AI slop and its \textit{mass producibility}.
They argue that a key feature of AI slop is that it is typically mass-produced \textit{with the intent of indiscriminately flooding social media feeds}\footnote{Consider, in fact, the tutorials widely available on TikTok on how to produce and monetize AI slop videos \cite{tiktokslop2026}.}, but this aspect of scale is entirely missing from the core example, which is meant to be an intuition pump for the article.

A second problem with the core example vis-à-vis the definition is the role of `communicative intent,' given the \textit{superficial competence} property of AI slop.
But the example where the child has a specific wish to see their own father perform particular movements clearly does not display a lack of communicative intent. 
Quite to the contrary, the child  wishes to express a very particular humorous message with the imagined scenario involving a person they know very well, which is precisely what makes the generated video all the more funny to the child. 
The example does not show the impersonal, mindless, intentionality-lacking characteristics which make AI slop ``lack communicative intent.''

The authors also claim that AI slop serves a \textit{social} function, but their core argument and example deal with the purely private consumption and preference satisfaction of one individual. 
The only way for Kommers et al.'s argument from the purely private to the social to work, is to argue that social utility is, by definition, nothing more than the sum of individual utility understood as simple preference satisfaction for individuals detached from their community and society. 
To be sure, such an argument, if made at all, would be very nihilistic.

Toward the argument for aesthetic value of AI slop, the analogy to `low' culture is not sufficient to establish the aesthetic value of AI slop.
For their indirect argument to work, the `low' art they reference should be sufficiently similar to AI slop; i.e., the `low' art should also be completely devoid of communicative intent and substance because if it is not, it also does not have explanatory value or justificatory power for AI slop. 
No such argument is provided, and in Section 4, we explain why this analogy might not work, through a discussion of the conditions that made the reappraisal of `low' culture possible.

\section{What Is Authentic Demand?}

We now turn to the external critique, starting with the discussion on \textit{demand}.
Kommers et al. write that ``People want more content than humans can supply''. The authors take the demand, that allegedly 
happens to exist, as 
natural starting point of the analysis. But demand doesn't emerge naturally, it is created. Big Tech companies design their (manipulative \cite{sax2021between, sax2021optimization}) interfaces to boost engagement 
in an attempt to manufacture ongoing demand for their services \cite{zuboff2023age, ienca2023artificial, de2025attention}. Some also point to ``a program of systematic desocialization'' \cite{herrmanInstagramReelsNew2026}, where users are trained to relate to others as content to be consumed constantly, rather than people to connect with. AI slop partly serves the need for constant consumption that this system creates, and calling that a social function unfairly legitimizes what is better understood as a harmful restructuring of how people relate to culture and to each other. 

So, even if it were true that people want more content than humans can supply, there is an important normative question begging to be asked: 
Should demand that is manufactured for the sole purpose of promoting the interests of Big Tech be treated as equally important as \textit{authentic} demand? Moreover, by treating the preferences of users as an unproblematic starting point one also makes end-users responsible for how they are using platforms, and, in the end, how they produce and consume AI slop. This is a dangerously one-sided argument that ignores the strategies of platform to \textit{create} this behavior; and potential demand in AI slop. In sum, the phenomenon of demand cannot be treated as an unproblematic starting point of AI slop analyses, but should itself be part of critical analyses.

\section{What Is the Aesthetic Value of AI Slop?}

Kommers et al. draw on the history of kitsch to argue that AI slop, like other `low' cultural forms initially dismissed by critics, may turn out to have aesthetic value. However, this analogy only works if kitsch (or any `low' culture) and AI slop are similar not just in how they are produced but in how they function culturally.
In fact, the observation that critics have sometimes been wrong about `low' culture would also support rehabilitating phenomena like spam and clickbait. 
The argument needs a positive account of why Al slop belongs to the class of things that were rehabilitated rather than the class that were not.

As \citet{greenbergAvantgardeKitsch1939} argues, kitsch is parasitic on `high' culture to derive its forms and standards.  
However, it is this relational quality that made  kitsch's later reappraisal possible.
\citet{calinescuFiveFacesModernity1987} documents that critics dismissed kitsch on the grounds that its value seemed \textit{merely} social---audiences consumed kitsch to chase social status, rather than for intrinsic aesthetic merit---but this dismissal undervalued the social dimensions of kitsch, i.e., that it spoke to the social aspirations of its audiences and generated shared interpretive frameworks and discourses around it \cite{geertzInterpretationCulturesSelected2009}.

AI slop, as defined, has a veneer of quality without aspiration toward it, meaning that the conditions which made the reappraisal of kitsch possible are undermined by slop.
First, kitsch was in dialogue with recognizable aesthetic standards, where slop is not, since it lacks communicative intent by definition.
Second, kitsch circulated as shared cultural objects, but hyper-personalized slop (e.g., the dancing father video) is designed for individual consumption, which may thwart the formation of shared interpretive frameworks.
Third, kitsch accumulated its cultural significance through sustained engagement from consumers and critics over time, but the zero-effort, mass-produced nature of AI slop work against this.
We can see these conditions at work when we consider other `low' cultures now viewed more favorably by critics.  
Fanfiction attracts scholarly attention because the conditions for reappraisal are present. 
Communities are built around shared aesthetic frameworks (genre conventions, shared canon, common tropes), sustained through reciprocal exchange, like gifting and feedback \cite{fieslerGiftLogicLabors2021, chengFeedbackExchangeOnline2022}.

There is one more important aspect about the analogy to `low' culture, which is that while kitsch was parasitic on high culture, it did not threaten to make cultural artifacts harder to create or sustain.
AI slop at scale, as \citet{simpsonEnshittificationCreativeInternet2025} document, degrades the infrastructures that creative communities depend on, and can crowd out human creative work \cite{warzelToolThatCrushes2025}.
Our argument here is that the analogy to `low' culture is not sufficient to establish the aesthetic value of AI slop, and in fact, a more urgent question pertains to how AI slop is impacting conditions (e.g., the infrastructures, communities, shared frameworks) through which aesthetic value is generated and sustained by people.

\section{The (Hidden) Costs of AI Slop}

According to the authors, one core argument for the appeal of AI slop is that ``it offers a near-zero cost path to achieving one’s goals.'' Much like their arguments based on the phenomenon of demand, here the authors also treat the phenomenon of cost as an unproblematic notion. They only focus on costs \textit{for the individual prompting} and only consider cost in a narrow economic sense. We present four considerations on the \textit{social} costs. These critical considerations also apply to generative AI more generally, which is unsurprising given the close connection between the phenomena of generative AI and AI slop. We do, however, think that there is a specific AI slop angle to all considerations. Built into the concept of AI slop is an element of quantity; it can be generated easily, quickly, in enormous quantities. Its ability to \textit{flood} online spaces helps explain specific risks and harms, making it a case of a difference in degree becoming a difference in kind. So, even though these considerations are also relevant to discussions on generative AI, they are especially relevant for the discussion of AI slop.

\textbf{Power}. Both Generative AI companies and social media platforms stimulate and incentivize the production and circulation of AI slop. 
As users, both as consumers and producers of AI slop, keep engaging with Big Tech companies, they help consolidate their market power \cite{birch2025undermining}. And nowadays with market power comes real political power, as the Trump regime has shown \cite{coeckelbergh2026technofascism}. Participating in the AI slop economy thus also has potential political consequences.

\textbf{Environment}. Creating AI slop is compute intensive. And as the authors explain, AI slop is especially attractive because it can be used to satisfy infinitely many idiosyncratic personal preferences with highly personalized videos. So, promoting the social value of AI slop comes with an environmental footprint \cite{crawford2021atlas, jegham2025hungry}. Ultimately, the normative question arises if the mass-production of AI slop justifies an increasing amount of environmental harm.

\textbf{Data Privacy}. AI slop can also come with privacy costs. In the authors' own example, a video is generated of an existing father in his underwear. It is telling that the example does not even consider the question whether this father wants this to happen to him. Only the immediate preference satisfaction of the prompter is considered. The example clearly shows issues around \textit{consent} of people who are shown in AI slop output \cite{golda2024privacy}.

\textbf{(Stochastic) Manipulation}. The ability to generate enormous amounts of highly specific content at almost no cost \textit{to the prompter} comes with a potentially massive \textit{collective} cost: AI slop for online 
campaigns that can manipulate political sentiments and possibly even elections \cite{ec2024tiktok}. 
Even if individual pieces of slop do not manipulate individual users, AI slop's ability to flood online spaces 
may have nontrivial impacts on collective behavior \cite{benn2022s}.

\section{Conclusion}

Kommers et al. \cite{kommers2025slop} argue that AI slop matters. We agree that AI slop is a phenomenon that deserves serious scrutiny, but in this short paper we explained why AI slop matters for reasons different than those presented by Kommers and colleagues. Where their argument depends on several problematically uncritical assumptions vis-à-vis the demand for -- and costs and consumption of -- AI slop, we argue that research on AI slop must engage with ethical and social science literature to substantiate claims about its societal function and aesthetic value. We call AI slop researchers to adjust or extend their research focus to the following research areas:

\textbf{Consumption of AI slop}. Research on AI slop should engage with -- normatively and empirically -- what constitutes AI slop demand. Social science researchers, for instance, could conduct qualitative focus groups or interview studies to sketch out rationales of why people are (not) consuming AI slop. Quantitative survey studies could provide statistical evidence in showcasing general consumption patterns and describe conditions under which AI slop is (not) consumed. These factors can include, among others, AI literacy, acts of resistance, authenticity perceptions, and risk-benefit trade-off perceptions. Normative and critical studies could inform theory building and specifically interrogate questions of public values and authenticity in AI slop consumption. 

\textbf{AI slop and art}. While we discussed arguments against the definition of aesthetic value brought forward by Kommers et al., it remains an open question in how far AI slop can create any form of aesthetic value. This line of research can be further explored by cultural and humanities scholars as well as artists. We propose that research should look into conditions under which collective sense-making and cultural values can be created around, or with the help of AI slop. 

\textbf{Regulatory interventions towards AI slop}. AI slop comes with a plethora of (hidden) societal costs. If we want to design effective regulatory interventions to curtail the harms of AI slop, those societal costs should be part of the regulatory discourse. Academic research on AI slop that builds on critical perspectives on, for instance, political economy, environmentalism, data privacy, and manipulation can inform that regulatory discourse. A richer understanding of the societal costs associated with AI slop will also help regulators explore the potential of important recent legislation like the European Union's AI Act, Digital Services Act, or the General Data Protection Regulation to address AI slop harms.

\section*{Author Contributions}
All authors contributed equally to this research.

\section*{Declaration of Interest}
The authors declare no conflict of interest.

\section*{Funding}
Marijn Sax was funded by the Dutch Research Council, grant number VI.Veni.221F.006. Kimon Kieslich was funded by the German Academic Exchange Service (DAAD) with funds from the Federal Ministry of Research, Technology and Space (BMFTR).

\bibliographystyle{ACM-Reference-Format}
\bibliography{references}
\end{document}